\documentclass[12pt]{article}
\usepackage{graphicx}
\usepackage{cite}
\usepackage{mathtools}
\usepackage{amssymb}
\usepackage{amsmath}
\usepackage{slashed}

\newcommand\pubnumber{NuPhys2018-Rosauro}

\def\Spain{$^a$ Departamento de F\'isica Te\'orica and Instituto de F\'{\i}sica Te\'orica, IFT-UAM/CSIC,\\
Universidad Aut\'onoma de Madrid, Cantoblanco, 28049, Madrid, Spain\\
$^b$ Department of Physics, School of Engineering Sciences, \\ KTH Royal Institute of Technology, AlbaNova University Center, \\ Roslagstullsbacken 21, SE--106 91 Stockholm, Sweden\\
$^c$ Institute for Particle Physics Phenomenology, 
Department of Physics,\\ Durham University, 
South Road, Durham DH1 3LE, United Kingdom}

\def\Title#1{\begin{center} {\Large #1 } \end{center}}
\def\Author#1{\begin{center}{ \sc #1} \end{center}}
\def\Address#1{\begin{center}{ \it #1} \end{center}}

\newcommand\pubblock{\rightline{\begin{tabular}{l} \pubnumber\\
         \end{tabular}}}
\newenvironment{Abstract}{\begin{quotation}  }{\end{quotation}}
\newenvironment{Presented}{\begin{quotation} \begin{center} 
             PRESENTED AT\end{center}\bigskip 
      \begin{center}\begin{large}}{\end{large}\end{center} \end{quotation}}

\def\beq{\begin{equation}}
\def\eeq#1{\label{#1}\end{equation}}
\def\eeqn{\end{equation}}

\def\beqa{\begin{eqnarray}}
\def\eeqa#1{\label{#1}\end{eqnarray}}
\def\eeqan{\end{eqnarray}}

\let\bar=\overbar

\def\Dslash{\not{\hbox{\kern-4pt $D$}}}
\def\dslash{\not{\hbox{\kern-2pt $\del$}}}

\def\msb{{\bar{\ssstyle M \kern -1pt S}}}

\begin{document}
\begin{titlepage}
\pubblock

\vfill
\Title{Neutrino-Dark Matter Portals}
\vfill
\Author{ M. Blennow,$^{a, b}$ E. Fern\'andez-Mart\'inez,$^a$ A. Olivares-Del Campo,$^c$\\
 S. Pascoli,$^c$ \textbf{S. Rosauro-Alcaraz},$^a$ A. V. Titov$^c$}
\Address{\Spain}
\vfill
\begin{Abstract}
The nature of dark matter is one of the open problems of the Standard Model of particle physics. Despite the great experimental efforts, we have not yet found a positive signal of its interactions with ordinary matter. One possible explanation would be that the dark matter particle is primarily coupled to another elusive particle, neutrinos. In this work we study this possibility with several realisations.
\end{Abstract}
\vfill
\begin{Presented}
NuPhys2018, Prospects in Neutrino Physics\\
Cavendish Conference Centre, London, UK, December 19--21, 2018
\end{Presented}
\vfill
\end{titlepage}
\def\thefootnote{\fnsymbol{footnote}}
\setcounter{footnote}{0}

\section{Introduction}
The unknown origin of neutrino masses together with the existence of the dark matter (DM) component of the Universe constitute our most significant experimental evidence for physics beyond the Standard Model (SM) and therefore the best windows to explore new physics. Neutrinos and DM also share their elusive nature, having very weak interactions with the other SM particles. A tantalising avenue of investigation is the possibility that a stronger connection between these two sectors exists. In this case, the best way to probe DM would be through the neutrino sector.



In this contribution, based on~\cite{Blennow:2019fhy} we will investigate some gauge-invariant SM extensions that lead to sizeable neutrino-DM interactions, exploring if neutrino probes could dominate our sensitivity to the dark sector. If DM does not participate in any of the SM gauge interactions, the natural expectation is that the strongest connection to DM will be via singlets of the SM gauge group. 
The neutrino portal includes the addition of singlet right-handed (RH) neutrinos $N_R$, which makes this option particularly appealing in connection to the evidence of neutrino masses and mixing.


\section{Coupling to the full lepton doublet}
In this section we will study the simplest scenario, in which DM couples directly to the full SM $SU(2)$ lepton doublet. We will adopt an effective field theory approach in order to avoid specifying the nature of the mediator. 
The Lagrangian describing the neutrino-DM interaction is thus given by
\begin{equation}
\mathcal{L} = \mathcal{L}_{SM}+\mathcal{L}_{DM}+\frac{c_{\alpha}}{\Lambda^2}\overline{\chi}\gamma_{\mu}\chi\overline{L_{\alpha}}\gamma^{\mu}L_{\alpha},
\label{Eq:LagBruteForce}
\end{equation}
with $\mathcal{L}_{DM}=\overline{\chi}\left(i\slashed{\partial}-m_{\chi}\right)\chi$, being $\chi$ a Dirac fermion DM particle.
From Eq.(\ref{Eq:LagBruteForce}) it is clear that DM couples as strongly to neutrinos as to charged leptons.

\begin{figure}
\centering
\includegraphics[width=5.5cm]{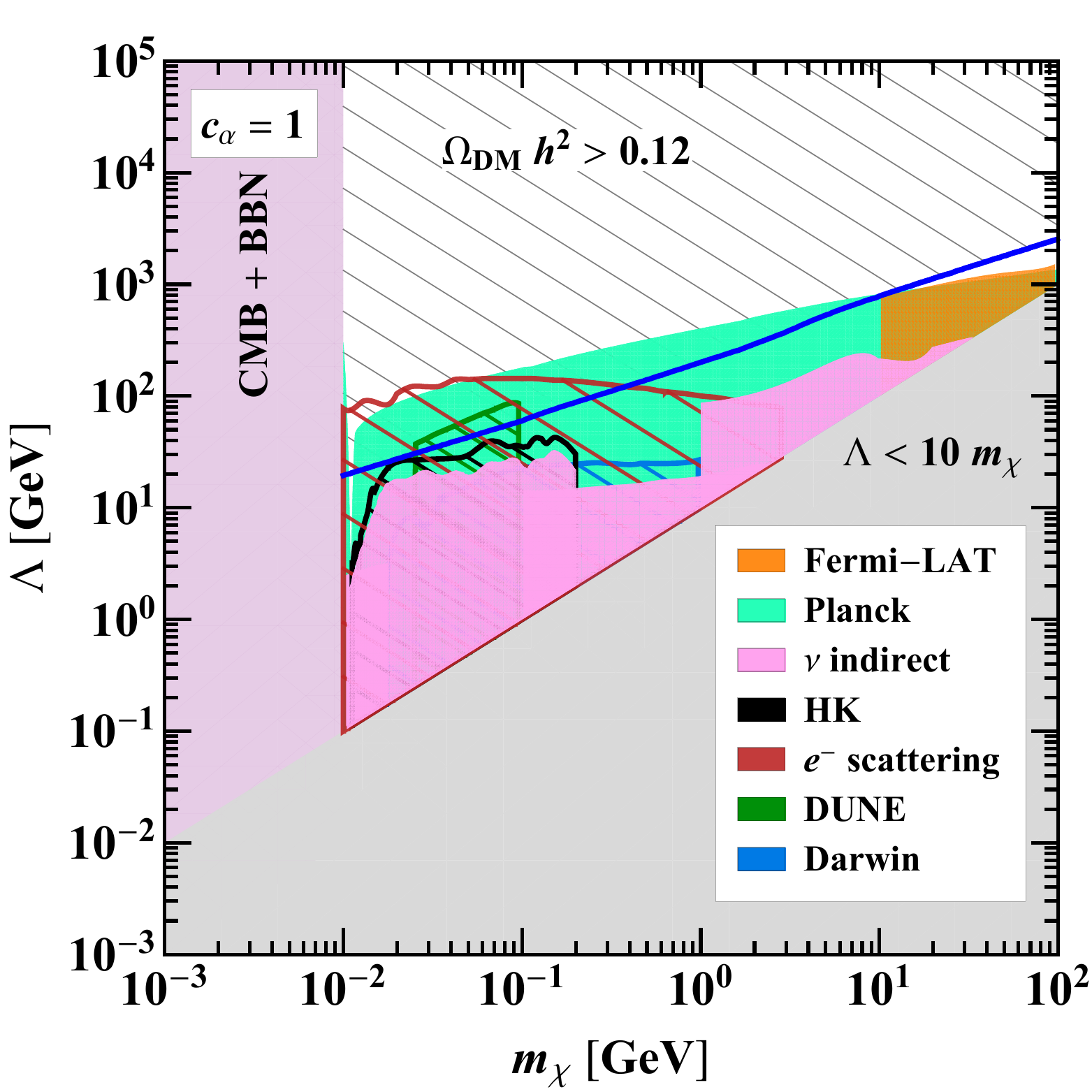}
\includegraphics[width=5.5cm]{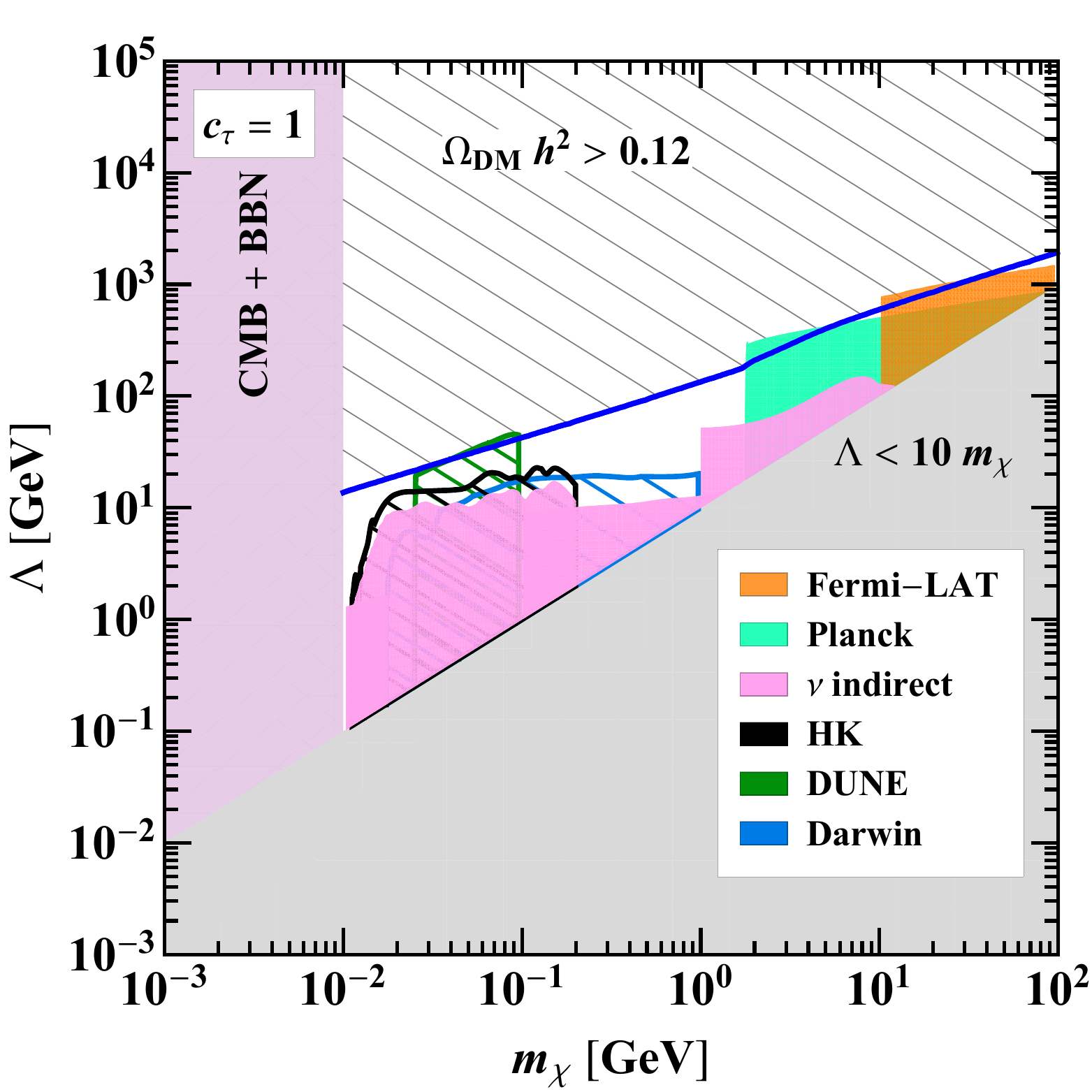}
\caption{Constraints on the DM mass $m_{\chi}$ and the new physics scale $\Lambda$. The left (right) panel corresponds to coupling to every lepton (only tau) doublets.  
}
\label{Fig:BruteForce}
\end{figure}


As can be seen in Fig.\ref{Fig:BruteForce}, whenever 
the annihilation to the corresponding charged lepton is possible, the constraints from Fermi-LAT~\cite{Ahnen:2016qkx} and the CMB~\cite{Aghanim:2018eyx} 
dominate, ruling out all the parameter space. Only if $m_{\chi}$ is smaller than the mass of the corresponding charged lepton neutrino experiments 
 control the sensitivity to these interactions. Future neutrino and DM experiments 
 could further explore these unconstrained regions as shown in the right panel of Fig.\ref{Fig:BruteForce}. For more details see Ref.~\cite{Blennow:2019fhy}.

\section{Coupling via the neutrino portal}
Given the results of the previous section, we will now explore  whether the neutrino portal option is able to lead to a rich DM-neutrino phenomenology without being in conflict with indirect searches involving charged leptons. In order to have sizeable mixing between the SM neutrinos and the extra $N_R$ we will attribute the smallness of neutrino masses to an approximate lepton number symmetry
. We will consider the addition of only one Dirac sterile neutrino that will serve as portal between the SM neutrinos and the DM. The Lagrangian of the model is given by
\begin{equation}
\mathcal{L}_N = \mathcal{L}_{SM}+\overline{N}\left(i\slashed{\partial}-m_N\right)N-\lambda_{\alpha}\overline{L_{\alpha}}\tilde{H}N_R.
\end{equation}
After electroweak symmetry breaking we get three masless neutrinos and a heavy one with mass $m_4 = \sqrt{m_N^2+\sum_{\alpha}v^2|\lambda_{\alpha}|^2}$. 
The amount of mixing between SM and the singlet neutrinos will be controlled by the size of $\theta_{\alpha}\equiv v\lambda_{\alpha}/m_N$
~\cite{Fernandez-Martinez:2016lgt}.


\subsection{Neutrino portal with a scalar mediator}
In this section we will assume that DM is composed of a new fermion, singlet under the SM gauge group, and that a new scalar mediates the neutrino-DM interactions. The Lagrangian of the model is given by
\begin{equation}
\begin{split}
\mathcal{L}=&\mathcal{L}_N+\mathcal{L}_{DM}+\partial_{\mu}S\partial^{\mu}S^*-y_L\overline{\chi}N_LS+h.c.-V(|S|^2,H^{\dagger}H),
\end{split}
\end{equation}
where $S$ is a complex scalar and $\chi$ is a Dirac fermion DM candidate. Note that the Lagrangian respects $U(1)_L$ lepton number and a global dark symmetry under which $\chi$ and $S$ have the same charge. In order to have DM stability we will assume $m_{\chi}<m_S$, where $m_S$ is the mass of the scalar.
The sensitivity to these interactions will be controlled by neutrino experiments, as the coupling to other SM particles appears at loop level.

\begin{figure}
\centering
\includegraphics[width=5.5cm]{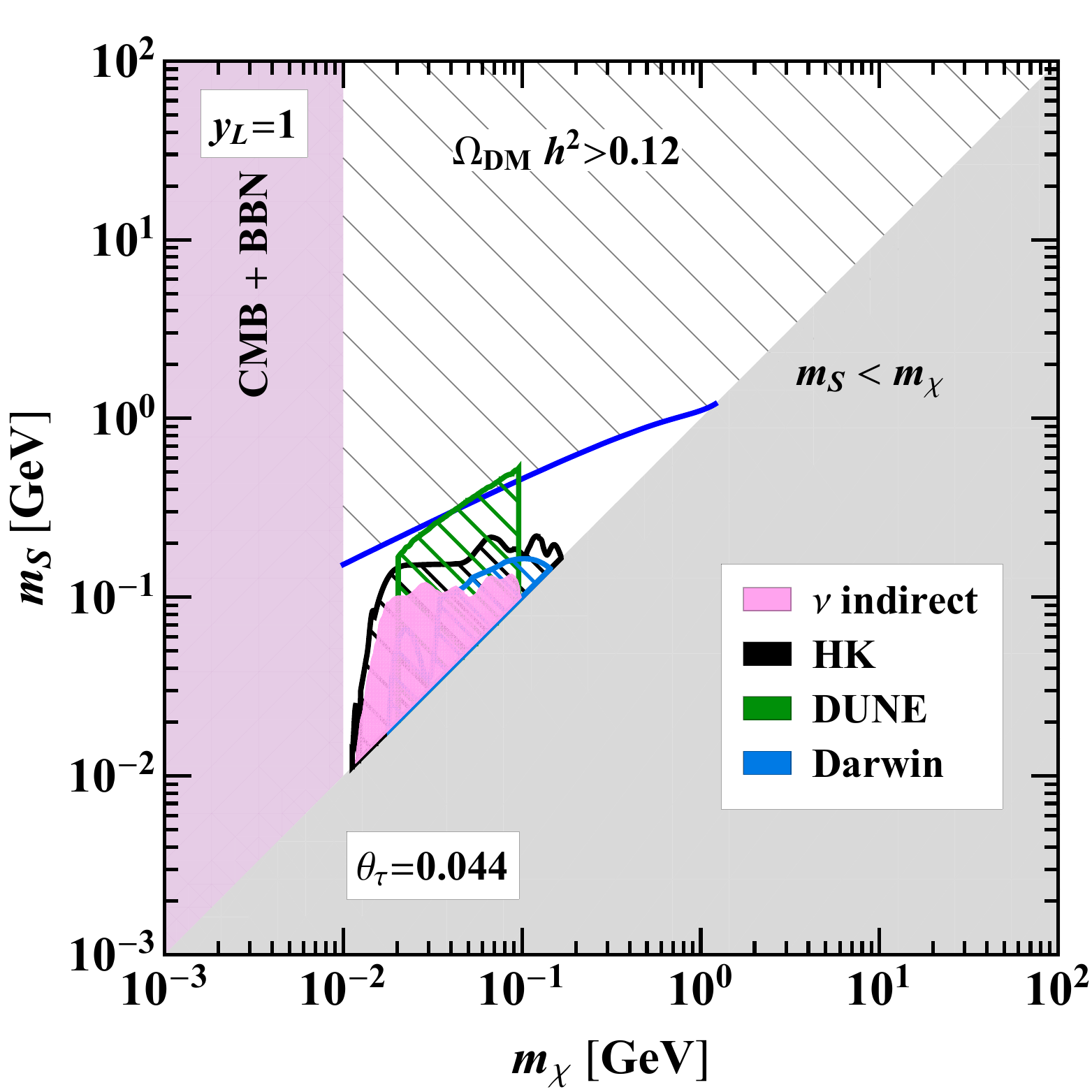}
\includegraphics[width=5.5cm]{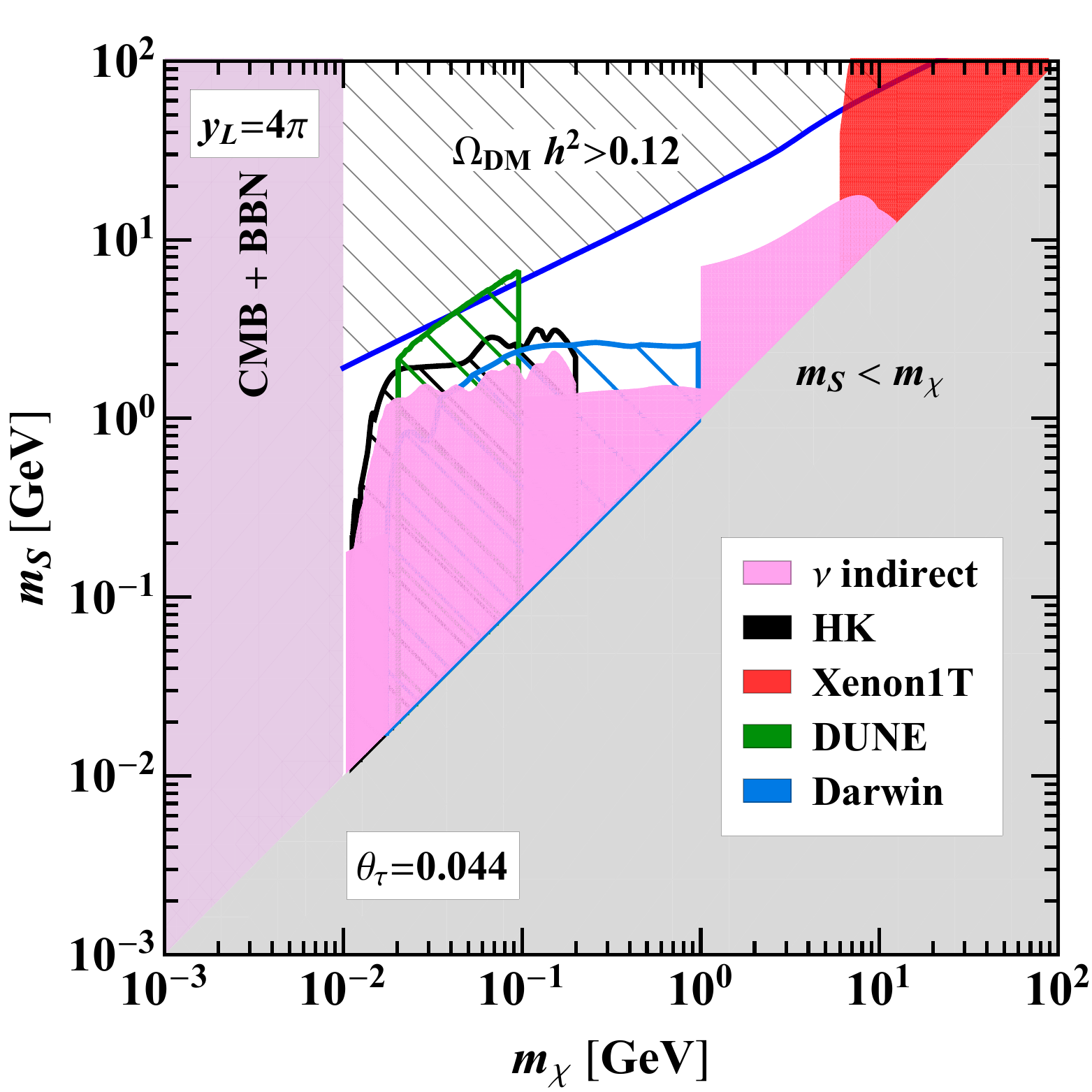}
\label{Fig:Scalar}
\caption{Constraints on the DM mass $m_{\chi}$ and the dark scalar mass $m_{S}$. We have fixed $\theta_{\tau}=0.044,\; \theta_{e,\mu}=0$. 
}
\end{figure}

Along the blue line in Fig.\ref{Fig:Scalar} the measured DM relic density~\cite{Aghanim:2018eyx} is obtained. Above this line the relic density is larger, leading to overclosure of the Universe.
Indirect searches for annihilation to neutrinos, together with direct detection bounds by XENON1T~\cite{Aprile:2018dbl}, are the only probes that constrain the allowed parameter space. The prospects to explore the remaining allowed regions are very promising (see Ref.~\cite{Blennow:2019fhy}). 

\subsection{Neutrino portal with a vector mediator}
In this second example, we will couple the Dirac DM fermion to a new massive vector boson. The Lagrangian of the model is given by
\begin{equation}
\begin{split}
\mathcal{L}=&\mathcal{L}_N+\mathcal{L}_{DM}-\frac{1}{4}Z'_{\mu\nu}Z'^{\mu\nu}+\frac{1}{2}m_{Z'}^2Z'_{\mu}Z'^{\mu}+g'\overline{\chi_R}\gamma^{\mu}\chi_RZ'_{\mu}+g' \overline{N_L}\gamma^{\mu}N_LZ'_{\mu},
\end{split}
\end{equation}
where $Z'$ is a new vector boson mediating the interaction between neutrinos and DM. This Lagrangian could describe a new spontaneously broken $U(1)'$ gauge symmetry. We will assume there is an additional conserved charge not shared between the neutrino and the DM preventing their mixing.
The phenomenology will be dominated by neutrino experiments, as the coupling to other SM particles appears at loop level, e.g., through kinetic mixing between the $Z$ and $Z'$ vector bosons~\cite{Blennow:2019fhy}.

\begin{figure}
\centering
\includegraphics[width=5.5cm]{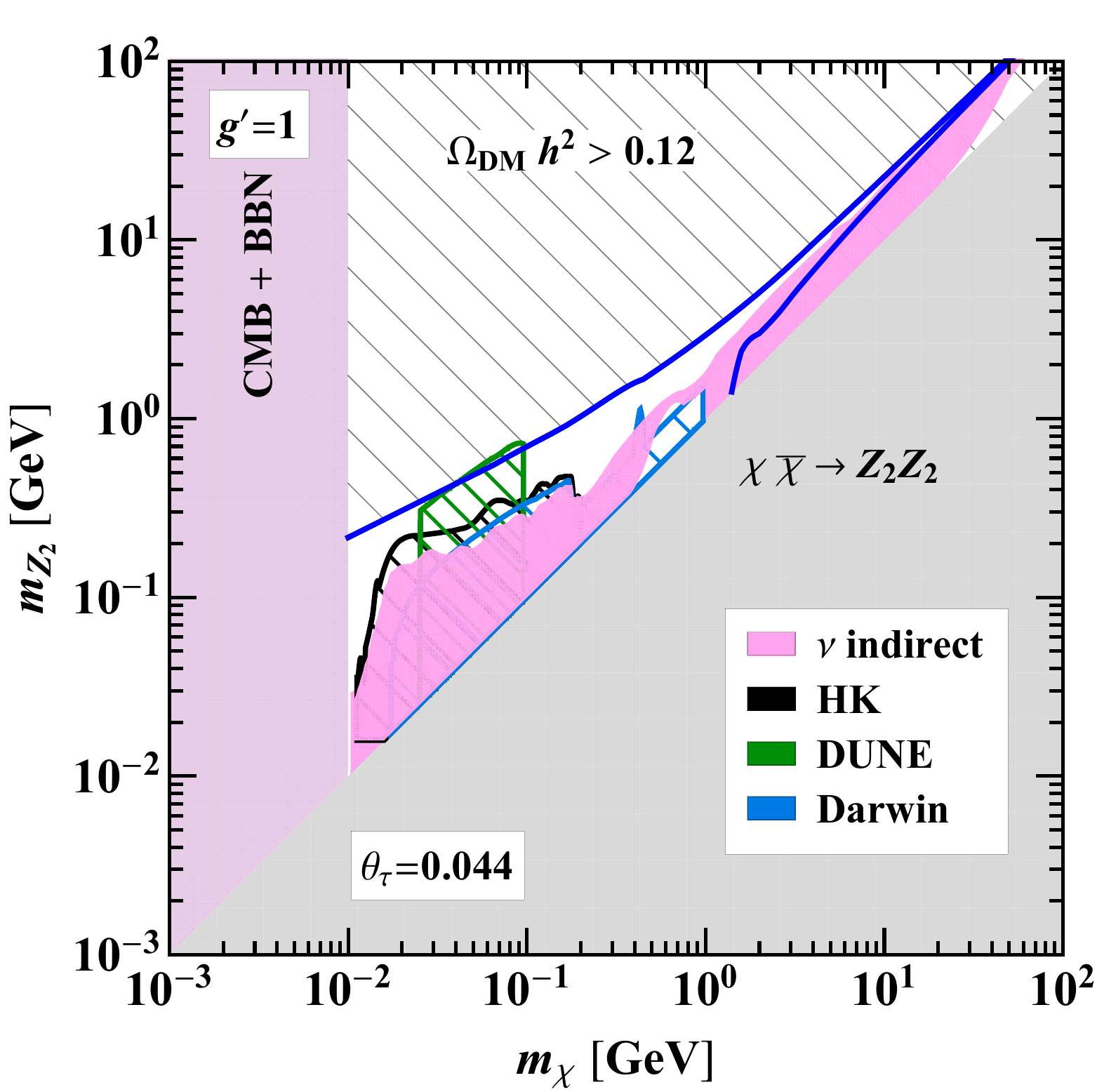}
\includegraphics[width=5.5cm]{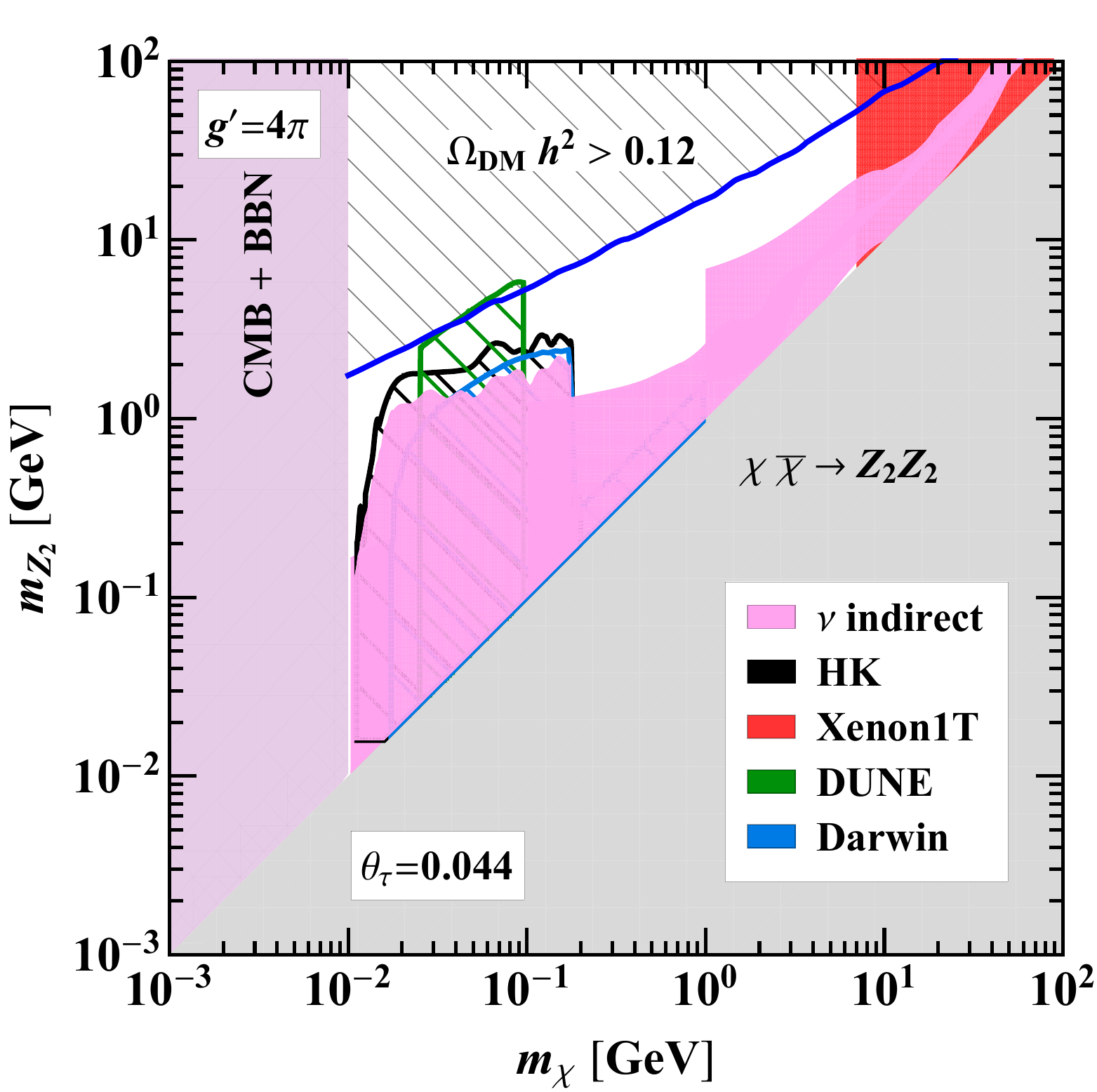}
\label{Fig:Vector}
\caption{Constraints on the DM mass $m_{\chi}$ and $m_{Z_2}$ for $\theta_{\tau}=0.044,\;\theta_{e,\mu}=0$. 
}
\end{figure}

In Fig.\ref{Fig:Vector} there are two branches where the DM relic density is obtained due to a resonant behaviour of the annihilation cross section to neutrinos. Large regions of parameter space are constrained through indirect detection at neutrino detectors
. 
Future neutrino experiments will further probe other regions of parameter space, with DUNE being able to probe the relic density target~\cite{Blennow:2019fhy}
.

\section{Conclusions}

We have explored whether a dominant neutrino-DM interaction is allowed in simple gauge-invariant models
. We first explored the simplest scenario, in which DM couples to the full lepton doublet. Whenever DM is heavier than the charged lepton(s) it couples to, the bounds from Fermi-LAT and CMB preclude DM-neutrino couplings sizeable enough to be probed.

We have then considered the option of the neutrino portal to DM, where DM couples directly to new heavy neutrinos, which are a natural addition to the SM to account for neutrino masses and mixing.
In the two realisations we explored we find that neutrino detectors place the most stringent and competitive bounds
. Future projects 
will be able to probe if the right-handed singlet fermions that can explain the origin of neutrino masses also represent our best window to the dark matter sector.

\bibliographystyle{JHEP}
\bibliography{NuPhys}


\end{document}